
\def\doublespace{\baselineskip=2\normalbaselineskip}
\doublespace
\def\Rf{\baselineskip=2\normalbaselineskip
\parindent=0pt \medskip\hangindent=3pc
\hangafter=1 }
\def\gtorder{\mathrel{\raise.3ex\hbox{$>$}\mkern-14mu
             \lower0.6ex\hbox{$\sim$}}}
\def\ltorder{\mathrel{\raise.3ex\hbox{$<$}\mkern-14mu
             \lower0.6ex\hbox{$\sim$}}}
\def\tcar{\futurelet\next\testnextcar}
\def\testnextcar{\ifhmode\ifcat\next.\else\ \fi\fi}

\def\tss{$\theta_{ss}$\tcar}
\def\epsd{$\epsilon ^D$\tcar}
\def\epsm{$\epsilon ^M$\tcar}
\def\tdr{$\theta^{D}_{R}$\tcar}
\def\tm{$\theta^M$\tcar}
\def\md{$m_D$\tcar}

\def\Sr{$S_R$\tcar}

\def\nuL{$\nu_L$\tcar}
\def\nuR{$\nu_R$\tcar}

\centerline{\bf  SEE-SAW ENHANCEMENT OF LEPTON MIXING}
\vskip 1cm
\centerline {\bf Alexei Yu. Smirnov}
\centerline{\it Institute for Advanced Study, Princeton, NJ 08540, USA}
\centerline{\it International Centre for Theoretical Physics,
34100 Trieste, Italy}
\centerline{\it Institute for Nuclear Research, 117312 Moscow, Russia}
\vskip 1cm
\centerline{\bf Abstract}

The see-saw mechanism of neutrino
mass generation may enhance   lepton mixing up to maximal
even if the Dirac mass matrices of leptons have structure
similar to that in the quark sector. Two sets of  conditions for such
an enhancement are found. The first one  includes
the see-saw generation of heavy Majorana masses for
right-handed neutrinos and a universality of
 Yukawa couplings which can follow from the unification
of neutrinos with new superheavy neutral leptons. The  second set
is related to
lepton number symmetry of the Yukawa interactions in
the Dirac basis of neutrinos. Models which realize these
conditions  have strong hierarchy or strong
degeneration of Majorana masses of the right-handed neutrinos.
\vskip 1cm

\noindent
{\bf I. Introduction}

The see-saw  mechanism of neutrino mass generation
 naturally relates a smallness of the neutrino
masses with  neutrality of neutrinos [1]. According to the see-saw
mechanism the following mass terms are introduced in the fermion
family basis:
$$
{\cal L}_m =
- \bar{\nu}_L m_D \nu_R - \nu_{R}^{T} C^{-1} {1 \over 2} M  \nu_R +
h.c.\ .
\eqno(1)
$$
Here \md is the Dirac mass matrix (Dirac sector) and
$M$ is the Majorana mass matrix  for
right-handed neutrino components  (Majorana sector). At
$M \gg m_{D} $ the terms (1) generate the Majorana mass matrix for left
(active) components ($\approx \nu_L$) [1]:
$$
m^{maj} = m_D M^{-1} m_{D}^{T}\ .
\eqno(2)
$$
An  attractive feature of this mechanism is that
the Dirac matrix, \md, can be similar (in scale and structure)
to that in the quark sector which is naturally implied by
 quark-lepton symmetry and Grand Unification. The difference
in scales of neutrino masses, and probably in lepton mixing,
follows from the structure of expression (2) and from the
Majorana mass matrix  which  has no
analogy in the quark sector.

 Mixing of leptons in the see-saw mechanism was widely discussed
before [2 - 14] (see [2] for review).
For $m_D ({\rm leptons}) \sim m_D$ $ ({\rm quarks})$
the  lepton mixing turns out to be typically
 of the same order of magnitude as quark
mixing [3 - 13], i.e. lepton mixing is relatively small at least
between the first two generations. This was illustrated, in particular,
 by Monte-Carlo
studies of different configurations of the Dirac and Majorana
mass matrices [8], although some textures of matrices result in
large mixing [4, 6, 8].

At the same time it is difficult to expect that lepton
mixing  coincides precisely with quark mixing.
Mixing is related to masses but the masses of
the charged leptons and down quarks from the same fermion generations
are different. Neutrino masses,
if they exist, are much smaller than the masses of up
quarks. The results of the gallium experiments [15] with
solar neutrinos show that Cabibbo mixing cannot explain
the solar neutrino problem.

Moreover,
there are several hints that lepton mixing
might be much larger than that in  the quark sector.
The solar neutrino problem [15, 16] can be solved
by long length vacuum oscillations
(``just-so" ) [17, 18] or  by resonant flavor
conversion, the MSW-effect [19, 20].
The former requires the  values of neutrino
mixing angles $\theta$ and masses squared difference $\Delta m^2$:
$\sin^2 2\theta = 0.85 - 1.0$,
$\Delta m^2 = (0.8 - 1.1) \cdot 10^{-10}~ {\rm eV}^2$ [18].
The latter picks up two regions of neutrino
parameters, one of which involves large  mixing angles:
$\sin^2 2\theta = 0.6 - 0.9$ at
$\Delta m^2 = (10^{-7} - 10^{-5})~ {\rm eV}^2$ [20].
The deficit of the muon neutrinos in the atmospheric neutrino flux
can be explained by $\nu_{\mu} - \nu_e$ oscillations with parameters :
$\sin^2 2\theta = 0.5 - 0.9$,
$\Delta m^2 = (10^{-3} - 10^{-2})~ {\rm eV}^2$ [21].

As follows from (2) the masses of light neutrinos in these
regions correspond  to Majorana masses of the
right-handed neutrinos of the order of ($10^{10} - 10^{12}$) GeV.
This is much smaller than a possible Grand Unification scale
which may testify for a strong hierarchical structure in the
Majorana mass sector. Such a hierarchy (as we will show) may result
in  enhancement of lepton mixing.

In this paper we assume  that the Dirac mass matrices in
the lepton sector are similar to those in the quark sector and find conditions
at which lepton mixing is enhanced by the see-saw mechanism itself.
The paper is organized as follows. In Sect. II, considering
the two-neutrino case, we formulate general
conditions for lepton-mixing enhancement. Two sets of conditions
are found. In Sect. III and IV
we consider two scenarios  which realize these conditions.
Sect. V contains a discussion and summary of results.
Some technical details
are explained in the Appendix.

\vskip 0.8cm
\noindent
{\bf II. See-saw angle. Mechanisms of enhancement.}

We will first explain the mechanism of enhancement for the
two-neutrino case and then generalize the results for three-neutrino mixing.

The mixing angle for two generations of leptons, $\theta^{lept}$,
can be written in the following form:
$$
\theta^{lept} = \theta^{D}_{L} - \theta^{l}_{L} + \theta_{ss}\ .
\eqno(3)
$$
The first two terms in the right-hand side of this
equation are the direct analogies
of  mixing angles in the quark sector: $\theta^{l}_{L}$ is the
angle of rotation of the left charged lepton components
which diagonalizes the mass matrix of charged leptons;
$\theta^{D}_{L}$ is the angle of rotation of the left neutrino
components, \nuL, which diagonalizes the  Dirac mass matrix
of neutrinos, \md  \ (see Appendix).
The last term in (3), \tss, is the additional
angle that specifies the effect of the see-saw mechanism itself.
If there are no Majorana mass terms ($M = 0$), then \tss $= 0$. We
will call \tss the see-saw angle.

Let us suggest that the Dirac matrix contribution to  $\theta^{lept}$
is of the order of quark mixing angle: $\theta^{D}_{L} - \theta^{l}_{L}
\ltorder \theta _{c} \approx 13^0$ ($\theta_c$ is the Cabibbo angle)
, and that
$\theta^{D}_{L}$ and  $\theta^{l}_{L}$  are separately
 also small $\approx \sqrt {m_e \over m_{\mu}}$. We will
find the conditions at which \tss
is appreciably larger than
  $ \theta^{D}_L$ and $\theta^{l}_{L}$, so that
$\theta^{lept} \sim  \theta_{ss}$  and the latter may be close to
$45^0$.

The angle \tss is determined by the properties of both the Majorana, $M$,
as well as of the Dirac, \md, mass matrices, and it is convenient to
write down \tss in terms of mixing angles and
mass hierarchies that characterize $M$ and \md.
Namely, let \tm and \tdr be the angles of rotation which  diagonalize
$M$ and \md correspondingly (see Appendix). Let us define the
mass hierarchies
of $M$ and \md as
$$
\epsilon^{D} \equiv {m_1 \over m_2}, \hskip 1cm
\epsilon^{M} \equiv {M_1 \over M_2}\ ,
\eqno(4)
$$
where $m_i$ and $M_i$ (i = 1, 2) are
the eigenvalues of \md and $M$ respectively.
Then  \tss  is determined from (see Appendix)
$$
\tan 2 \theta_{ss} =
- 2\tan (\theta^M - \theta^{D}_{R}) \cdot
{
\epsilon^D \cdot (1 - \epsilon^M)
\over
\tan ^2 (\theta^M - \theta^{D}_{R}) + \epsilon^M - \delta
}\ ,
\eqno(5)
$$
where  $\delta$ is a small term of second order in the Dirac mass
hierarchy:
$$
\delta \equiv (\epsilon^D )^2 \left[ 1 + \epsilon^M
\tan ^2 (\theta^M - \theta^{D}_{R}) \right]\ .
\eqno(6)
$$
Two remarks are in order. As follows from (5, 6), the angle \tss
is precisely zero at \tm = \tdr \ (i.e. when the rotations
of right-handed components that diagonalize $M$ and \md
are the same), or at \epsm = 1, (when the masses
of Majorana sector are degenerate).
The angle \tss is proportional to the mass hierarchy
in the Dirac sector (rather than $\sqrt {\epsilon ^D}$),
and since \epsd is small one obtains
typically small values of \tss, unless the denominator
in the RHS of Eq.(5) is strongly suppressed.

Let us consider the conditions for \tss enhancement.
According to (5) there are two possibilities: (i)
all terms in the denominator of the RHS of the Eq.(5) are very small,
(ii) there is a strong cancellation  among two first terms.

(i) The first case corresponds to
{\it a strong mass hierarchy in the Majorana sector}. Indeed, according to
Eq.(5), one obtains $\tan 2 \theta_{ss} \gtorder 1$, when
$$
\tan (\theta ^{D}_{R} - \theta^M) \ltorder \epsilon^D\ ,
\eqno(7)
$$
and
$$
\epsilon^M \ltorder (\epsilon^D )^2\ .
\eqno(8)
$$
Note, if \epsm $> 0$, then the maximal value of \tss,
$$
\tan 2 \theta_{ss}^{max} \approx
- {
\epsilon^D
\over
\sqrt {\epsilon^M - (\epsilon ^D) ^2}
}\ ,
\eqno(9)
$$
is achieved at
$\tan (\theta ^{D}_{R} - \theta^M) = \sqrt{\epsilon^M - (\epsilon ^D) ^2}$.
Consequently, the equality  $\tan 2\theta_{ss}^{max} \approx 1$ implies
$\epsilon^M \sim (\epsilon^D )^2$; the mass hierarchy in
the Majorana sector should be of the order of the Dirac mass
hierarchy squared.

The condition  (7) means that the angles \tm and \tdr
are close to each other:
since  $\theta ^{D}_{R} \sim \sqrt {\epsilon^D}$, we get from (7)
$\theta ^{D}_{R} - \theta^M \ll \theta ^{D}_{R}$.

(ii). The condition of {\it strong cancellation in the denominator}
of the RHS of Eq.(5) reads
$$
\tan ^2 (\theta^M - \theta ^{D}_{R}) \approx - \epsilon^M\  .
\eqno(10)
$$
Now a strong mass hierarchy is not needed and, moreover ( as we will show),
 an approximate mass degeneration in the Majorana sector naturally
results in cancellation.
If the equality in Eq.(10) is exact, then according to (5)
$$
\tan 2 \theta_{ss} \approx
{
2\sqrt {|\epsilon^M |}
\over
\epsilon ^D
}\ .
\eqno(11)
$$
Consequently,  at \epsm $\approx$ \epsd, one has
$\tan 2 \theta_{ss} \sim
1/ \sqrt {\epsilon^D} \gg 1$; i.e. the angle \tss can be  close to the maximal
mixing value. The weaker the hierarchy \epsm, the larger
\tss and, consequently, the larger the lepton mixing, in contrast to
the previous case.

Since \tm is determined by  the Majorana matrix,
$M$, whereas \tdr is determined by the Dirac matrix, \md, the
equality (10) implies, in general, the relation between the
structures of these matrices. It is instructive to  express this relation
in the following form. Let us define the {\it Dirac basis} of
neutrinos,  $\nu^D_R$, as the basis
of the neutrino states in which \tdr = 0, i.e. in which
the Dirac mass matrix, \md, is diagonal. Then the condition
(10) means that in the Dirac basis  the Majorana mass matrix
should satisfy the following condition:
$$
\tan ^2 \theta^M \approx - \epsilon^M, \hskip 1cm (\theta^D _R = 0)\ .
\eqno(12)
$$
In turn, Eq. (12) implies the following inequality of
the matrix elements of $M$:
$$
M_{11} \cdot M_{22} \ll M_{12}^2\ .
\eqno(13)
$$
If $M_{11} \cdot M_{22} = 0$, the equality (12) is exact.

Let us comment on some special cases.

1) $M_{11} = 0$, $M_{22} \not = 0$. (Inequality
$M_{22} \gg M_{12}$ corresponds to
the Fritzsch ansatz [22] for
$M$).  One gets
near to maximal mixing according to Eq.~(11), and  an inverse mass hierarchy
in the light neutrino sector.

2) $M_{11} \not = 0$, $M_{22} = 0$. These conditions result in
an inverse mass hierarchy of the heavy
Majorana neutrinos:  \epsm $>$ 1, and in an inequality
$\tan ^2 \theta^M = -1/ \epsilon^M$.  Therefore,  a strong
cancellation will take place only at strong mass degeneration,
\epsm $\approx$ 1
($M_{11} \ll M_{12}$). In this case, the expression for
the angle \tss can be written as
$$
\tan 2 \theta_{ss} \approx
{
2\epsilon ^D
\over
1 - |\epsilon ^M|} = - {2\epsilon ^D M
\over \Delta M
}\ ,
\eqno(14)
$$
where $\Delta M$ is the difference in absolute values of Majorana masses.
(At $M_{11} = M_{22} = 0$ there are no mass splitting and no mixing;
the light components form a ZKM- neutrino.)

3) In  the more natural case, $M_{11} \sim M_{22} \ll M_{12}$,
one can get both mixing enhancement and a normal
mass hierarchy of light neutrinos (see Sect. IV).

Mixing is enhanced also if
 the condition (12) is fulfilled at nonzero
\tdr, but this angle should be small. Indeed substituting (12)
in (5) we get:
$$
\tan 2 \theta _{ss} \approx {\epsilon ^D \over \tan \theta ^{D}_{R}}\ ,
$$
i.e. for $\tan 2 \theta_{ss}$ to be of the order of 1, one needs
$\tan \theta ^{D}_{R} \leq \epsilon^{D} \approx 10^{-2}$,
which is much smaller than the typical rotation angle in the Dirac sector.
Consequently, a strong non-diagonality (13) of the Majorana mass matrix of
right-handed
neutrinos alone is not enough to get strong enhancement of mixing.

The above conditions can be immediately generalized for
 three-neutrino mixing. In the first case one should suggest
a strong mass hierarchy in the Majorana sector:
$M_i \propto (m_i^D)^2$ and that
the Majorana and the Dirac mass matrices are diagonalized by
approximately the same transformations  of
right-handed neutrino components $S_R$, (\Sr $\equiv S(\theta_{R}^{D})$ for
two neutrinos). In the second case, at least some of
the elements
of the Majorana mass matrix in the {\it Dirac basis of neutrinos}
should satisfy the condition $M_{ij}^2 \gg |M_{ii} \cdot M_{jj}|$.
Note that  such a non-diagonality of the Majorana mass matrix is
a generic feature of models where neutrino components have
nonzero lepton numbers and a symmetry related to this lepton
number is preserved (or approximately
preserved)  by the Yukawa couplings  and possible bare
mass terms.

\vskip 0.8cm

\noindent
{\bf III.  Enhancement of mixing in the case of strong mass
hierarchy\break\hglue30pt in the Majorana sector.}

Both conditions of the enhancement---strong
hierarchy in the Majorana sector (8), and approximate equality
of the rotation angles (7), are satisfied simultaneously if
$$
M \sim {1 \over \mu}  \cdot m_{D}^{T}  m_{D} \sim
{1 \over \mu} \cdot S_R (m_{D}^{diag})^2 S_{R}^{T}\ .
\eqno(15)
$$
Here $\mu$ is some mass parameter and \Sr
is the transformation  which diagonalizes  \md.
(Strictly speaking,
the last equality in (15) is true for the CP-conserving
\md). In case of exact equality (15), the matrix $M$ is also diagonalized
by \Sr,  i.e. \tm = \tdr, and \epsm = (\epsd)$^2$
for the two-neutrino mixing. In fact,
the equality (15) should be slightly broken because at
\tm = \tdr one gets  \tss $\equiv 0$.

The relation (15) can be obtained if the right-handed neutrino
components acquire Majorana masses  via the see-saw
mechanism too. This implies the existence of new heavy
neutral leptons and the ``cascade" , or extended see-saw
(see e.g. [6], where the extended see-saw was used for
another purpose). The first (high-mass scale)
see-saw induced by
the interactions of \nuR with  new leptons gives the Majorana
masses to \nuR 's, and the second one generates the masses to
the light neutrinos.

Let us introduce three (one for each fermion family)
neutral leptons $N_L = (N_{1L}$, $N_{2L}$, $N_{3L})$---singlets of
the electroweak symmetry---with
bare Majorana masses, $m_{N}$, at some large scale
(e.g. $m_N$ can coincide with  Grand Unification scale,
$M_{GU}$).
Let us introduce also
the scalar field, $\sigma$, singlet of $SU(2) \times U(1)$, which acquires
a vacuum expectation value $\sigma_0 \gtorder 10^{12}$ GeV.
Then the Yukawa interactions and the bare mass terms of the
model are
$$
\bar{\nu}_{L} h_{\nu} \nu_R \phi +
\bar{N}_L h_N \nu_R \sigma + N^{T}_{L} C^{-1} {1 \over 2} m_{N} N_L +
h.c.\ .
\eqno(16)
$$
Here $\phi$ is the usual Higgs doublet, and
$h_{\nu}$ and $h_{N}$ are the matrices of the Yukawa couplings. We
suggest that the bare Majorana masses of \nuR 's are small or
forbidden by some symmetry $G$. For example, one can
introduce the global $G = U(1)$  and prescribe  the
following G-charges: G(\nuR) = G(\nuL) = G($\sigma$) = 1,
G($N_L$) =  G($\phi$) = 0. Another possibility is that
the \nuR enters  the  doublet of $SU(2)_R$
 symmetry which is broken at the scale $V_R \ll M_{GU}$.
In this case  \nuR may acquire a mass
$\ll V_R$ [1].

To reproduce the relation (15) one should suggest that
the matrices of the Yukawa interactions, $h_{\nu}$ and $h_{N}$, are
correlated. The simplest possibility, $h_{\nu} = h_{N}$,
means universality of Yukawa interactions:
the couplings of \nuR with \nuL and $N_L$ are the same.
Such a universality implies unification; it can be realized
if \nuL and $N_L$ as well as $\phi$ and $\sigma$ enter the same
fermion and scalar
multiplets correspondingly. In fact, the equality of $h_{\nu}$ and  $h_{N}$
is not necessary;
the enhancement takes place if the matrices of
the Yukawa couplings have the following form:
$$
h_{\nu} \equiv S_{L} {m_{D}^{diag} \over v} S_{R}^{T}, \hskip 1cm
h_{N} = S_{N} {m_{D}^{diag} \over v} S_{R}^{T}\ .
\eqno(17)
$$
Here $v$ is the vacuum expectation value of $\phi$ and
$S_{N}$ is some almost-arbitrary matrix which may
be related to  features
of $N_{L}$ interactions.
 In particular, it can be equal to  $I$ or $S_{L}$.

When the fields $\phi$ and $\sigma$ acquire nonzero vacuum
expectation values, neutrino masses are generated and the
mass matrix in the block basis ($\nu_L$ , $\nu ^{c}_{L}$, $N_L$)
can be written as
$$
\pmatrix{
0 & m_D & 0 \cr
m_{D}^{T} & 0 & {\sigma _0 \over v} m_{D}^{T} S_{N}^{T} \cr
0 & {\sigma _0 \over v} S_{N} m_{D}  & m_N \cr
}\ .
\eqno(18)
$$
It gives the Majorana mass matrix for right-handed neutrino
components:
$$
M = - \left({\sigma _0 \over v}\right)^2 m_{D}^{T} S_{N}^{T} m_N ^{-1}
S_N m_D\ ,
\eqno(19)
$$
and then   the Majorana mass matrix for
light components:
$$
m^{maj} = \left({v \over \sigma_0}\right)^2 S_N ^T m_N S_N
\eqno(20)
$$
(for real $S_N$).
As follows from Eq. (20), at $S_N = I$,  the matrix $m^{maj}$
is proportional to the mass matrix of superheavy leptons,
$N_L$. We may suggest that the Yukawa couplings
or/and  bare mass terms at the highest mass scale have no
hierarchy; all elements of the matrix $m_N$ are of
the same order.
The matrix $S_N$ with small nondiagonal elements
gives only small corrections to the above picture.

Note that at $\sigma _0 \sim m_N \sim 10^{16}$ GeV, the
typical mass scale for the lightest components
is about $m \sim v^2 m_N / \sigma_0 ^2 \approx 10^{-2}$ eV.
Therefore a small
spread of parameters in $m_N$ allows us to explain
 the scales of both the solar  ($ m \sim 0.3 \cdot 10^{-2}$ eV)
and atmospheric ($m \sim (3 - 10) \cdot 10^{-2} {\rm eV}^2$) neutrino
problems. The Majorana masses of right-handed neutrinos are naturally in the
intermediate mass scale region ($10^{10} - 10^{14}$) GeV.

Two remarks are in order.
A Goldstone boson which could appear due to spontaneous
violation of G symmetry by the vacuum
expectation of $\sigma$ has negligibly small
interactions with active neutrinos. Moreover,
it can acquire nonzero mass due to explicit violation
of this symmetry at low scales.
Additional scalars can be introduced to
generate $m_N$ spontaneously.

The above mixing enhancement
based on the extended  see-saw mechanism and
the universality of the Yukawa couplings allows
us to systematically compensate for the smallness
in the lepton mixing related to the hierarchical structure of \md.
 In the extreme  case the lepton mixing is
defined by the structure of the mass matrix
of superheavy leptons and does not depend
on the Dirac matrix at all.

\vskip 0.8cm

\noindent
{\bf IV. Enhancement of mixing due to lepton symmetry
in the Dirac \break
\phantom{fghy}basis. Degeneration of heavy Majorana masses.}.

The enhancement of mixing takes place  due to the cancellation
in the denominator of the RHS of Eq.(5). The condition of cancellation
  (the relation (12) for the
Majorana mass matrix in Dirac basis of neutrinos) implies
a correlation between structures of the Dirac and the Majorana matrices.
Possible deviation from this correlation quantified by the
angle \tdr should be small. These features again imply
the universality of the Yukawa couplings. The neutrino family states,
at least \nuR,
should enter both
the Yukawa interactions and the bare mass terms in the same combinations, which
coincide with eigenstates of the Dirac mass matrix:
$ S_R ^{T} \nu_R \equiv \nu^D _R $. However, in contrast with
the previous case, the couplings of $\nu^D _R$ need not be proportional
to $m^{diag}_D /v$.
To naturally satisfy the condition (12),
the Yukawa interaction of $\nu^D_R$ should obey definite
symmetry.

It is possible to formulate these conditions in
another form. Usually lepton symmetry  and lepton numbers are
introduced in the Dirac basis of the {\it charged leptons}
(i.e. in the basis where
the mass matrix of charged leptons is diagonal; see e.g. [7]
in context of the see-saw mechanism). Lepton number
symmetry introduced in such a way is violated by
neutrino mass terms. In contrast, one can  impose the lepton symmetry
in the {\it neutrino} Dirac basis and suggest that this symmetry is
violated by mass terms of charged leptons.
These two cases are physically equivalent if there are only
Dirac masses, but they give different results in
presence of the Majorana mass terms. In the
neutrino Dirac basis, the lepton mixing matrix can be
written as
$$
S^{lept} = (S_L^l)^+ \cdot S_{ss}\ ,
\eqno(21)
$$
where  $S_L^l$ is the transformation of the
left components which diagonalizes the charged lepton mass matrix and
$S_{ss}$ is the see-saw  transformation  which diagonalizes
the mass matrix for light components in the Dirac basis (see Appendix).
In the two-neutrino case, Eq. (21) corresponds to Eq. (3)
at $\theta_L^D = 0$; by definition,
$S_L^l $ relates the Dirac and the flavor bases:
$\nu_f = S_L^l \nu_D$; $S_L^l$ is similar to the
Cabibbo-Kobayashi-Maskawa matrix of the quark mixing.

Consider neutrino mixing in the Dirac basis, $\nu_D$.
Let us suppose that the Yukawa interactions
as well as the bare mass terms obey  a global
G = U(1) symmetry. We prescribe zero G-charge for
the usual Higgs doublet, $G(\phi) = 0$, and different G-charges
for different pairs of left and right-handed neutrino components:
$G(\nu_{i L}) = G(\nu_{i R}) = G_i$, $G_i \not = G_j$
(i, j  = 1, 2, 3). Then the  Yukawa couplings of neutrinos with $\phi$ and
consequently the Dirac mass matrix are diagonal (as is demanded
by definition of the Dirac basis).
We will also introduce a scalar field $\sigma$ with
nonzero charge $G_{\sigma}$, which acquires nonzero vacuum
expectation value $\sigma_0$ and thus breaks G-symmetry.
Now Yukawa interactions and bare mass terms of the model can be
written as:
$$
\bar{\nu}_L {m_D ^{diag} \over v} \nu_R \phi +
            \nu_R ^T C^{-1} {1 \over 2} h \nu_R \sigma +
\nu_R ^T C^{-1} {1 \over 2} M_b  \nu_R\  + h.c.\ .
\eqno(22)
$$
Matrices of the Yukawa couplings, $h$, and bare masses, $M_b$,
are determined by the G-charge prescription and in general are
nondiagonal.
The neutrino mass matrix
which results from the  spontaneous breaking of the electroweak and
G- symmetries can be written in the basis ($\nu_L ^c , \nu_R$) as
$$\pmatrix{
0 & m_D ^{diag}  \cr
m_D ^{diag} & M \cr
}\ .
\eqno(23)
$$
The matrix of right-handed neutrino components, $M$, should
satisfy the cancellation
condition (see Eqs. (12) and (13) in two neutrino case).
According to the see-saw mechanism the mass matrix for light
neutrinos is
$$
m_{ss} = - m_D ^{diag} M ^{-1} m_D ^{diag}\ .
\eqno(24)
$$
Depending on $G_{\sigma}$ and on the charge prescription
for $\nu_{R} ^{D}$: $G_{\nu}  \equiv (G_1, G_2, G_3)$,
one gets different realizations of $M$ needed for
mixing enhancement. Let us
comment on the simplest possibilities.

1) For the charge prescription $G_{\nu} = (1, 0,-1)$, $G_{\sigma} = 2$
 the Majorana mass matrix is equal to:
$$
M =
\pmatrix{
h_{11} \sigma_0 & 0 & M_{13} \cr
0 & M_{22}& 0 \cr
M_{13} & 0 & h_{33} \sigma_0 \cr
}\ .
\eqno(25)
$$
The state $\nu_{2R}$ decouples and the task is reduced to the
two-neutrino case. At $h_{11} \sigma_0 \sim h_{33} \sigma_0 \ll M_{13}$,
one naturally  obtains a direct mass  hierarchy in the light sector and
a strong cancellation
in the denominator.
The see-saw transformation diagonalizing
matrix (24) is
$$
S_{ss} =
\pmatrix{
\cos \theta_{ss} & 0 & \sin \theta _{ss} \cr
0 & 1 & 0 \cr
- \sin \theta _{ss} & 0 & \cos \theta_{ss}\cr
}\ ,
$$
where according to (14)
$$
\tan 2 \theta_{ss} = {2 \epsilon^D \cdot M_{13} \over
[(\epsilon^D)^2 h_{33} - h_{11}] \sigma_0} \approx
{2 \epsilon^D M_{13} \over h_{11} \sigma_0}\ .
\eqno(26)
$$
Here \epsd $\equiv  m_1/m_3$.
The $\nu_e - \nu_{\tau}$
mixing turns out to be enhanced, whereas
$\nu_e - \nu_{\mu}$  and $\nu_{\mu} - \nu_{\tau}$ mixings
 induced only by Dirac matrices are small.

For the charge prescription
$G_{\nu} = (1, -1, 0)$, $G_{\sigma} = 2$ one  gets similarly
an enhancement of 1 - 2 and,  consequently, $\nu_e - \nu_{\mu}$ -
mixing. For $G_{\nu} = (0, 1, -1)$, $\nu_{\mu} - \nu_{\tau}$ mixing
is enhanced.

2) For $G_{\nu} = (1, 0,-1)$, $G_{\sigma} = 1$ the Majorana
mass matrix
has a form
$$
M =
\pmatrix{
0 & h_{12} \sigma_0 & M_{13} \cr
h_{12} \sigma_0 & M_{22} & h_{23} \sigma_0 \cr
M_{13} & h_{23} \sigma_0 & 0 \cr
}\ .
\eqno(27)
$$
A straightforward calculation of the $m_{ss}$ according to
(24) gives that at $h_{23} \sigma_0 \gg M_{13}$ the
enhancement
of 1 - 2 ($\nu_e - \nu_{\mu}$) mixing takes place:
$$
\tan 2 \theta _{ss} \approx
{
2  \epsilon ^D\cdot {M_{13} \over h_{23} \sigma_0}
\over
\left(\epsilon ^D \right)^2 - \left({M_{13} \over h_{23} \sigma_0}\right)^2
}\ ,
\eqno(28)
$$
where $\epsilon^D \equiv {m_1 \over m_2}$.

3) For $G_{\nu} = (0, 1, 2)$, $G_{\sigma} = 2$, the Majorana
mass matrix is

$$M = \pmatrix{
M_{11} & 0 & h_{13} \sigma_0  \cr
0 & h_{22} \sigma_0 & 0 \cr
h_{13} \sigma_0 & 0 & 0\cr
}\ .
\eqno(29)
$$
The state with charge $G = 1$ decouples and the
task is reduced again to the two-neutrino case.
In contrast to version  1), here
 mixing is induced by the Yukawa couplings with
$\sigma$. The enhancement of mixing implies
$h_{13}  \sigma_0 \gg M_{11}$ and according to (14) one has
$$
\tan 2 \theta_{ss} = 2 \epsilon ^D _{13} \cdot
{h_{13} \sigma_0 \over M_{11}}\ .
\eqno(30)
$$
The opposite order of charges
$G_{\nu} = (2, 1, 0)$
gives strong enhancement of 1 - 3 mixing, but an
inverse mass hierarchy of light neutrinos.
In this case, large mixing angles,  i.e. $\sin^2 2\theta > 0.6$,
are disfavored by SN1987A data [23].

Changing  the neutrino component  which has charge G = 1
and,  consequently, decouples, one can
enhance mixings of other pairs of neutrino states.

4). For $G_{\nu} = (2, 1, 0)$, $G_{\sigma} = 1$, the Majorana mass
matrix,
$$
M =
\pmatrix{
0 & h_{12} \sigma_0 & 0 \cr
h_{12} \sigma_0 & 0 & h_{23} \sigma_0 \cr
0 & h_{23} \sigma_0 & M_{33} \cr
}\ ,
\eqno(31)
$$
results in strong enhancement
of 1 - 2 ($\nu_e - \nu_{\mu}$)
mixing at $M_{33} \gtorder h_{12} \sigma_0$:
$$
\tan 2 \theta_{ss} = - {2 \over \epsilon^D _{12}}
{M_{33} h_{12} \over \sigma _0 h_{23}}\ ,
\eqno(32)
$$
and in  an inverse
mass hierarchy of light neutrinos.
At $h_{12}/h_{23} \ltorder \epsilon^D$ one can get simultaneously the
(1 - 3) -  mixing enhancement:
$$
\tan 2 \theta _{ss} \approx
{
2  \epsilon ^D\cdot {h_{23} \over h_{12}}
\over
1 - \left(\epsilon ^D \right)^2 \left({h_{23} \over h_{12}}\right)^2
}\ .
\eqno(33)
$$
For $G_{\nu} = (0, 1, 2)$, and $G_{\sigma} = 1$,
$\nu_{\mu} - \nu_{\tau}$- mixing is enhanced at $M \gtorder h \sigma_0$.

All elements of the Majorana mass matrix $M$ can
be generated spontaneously by
introducing additional scalar bosons, so that the hierarchy
of masses in the Majorana sector is related to the hierarchy
of the vacuum expectation values  of different scalars.
Moreover, the  extended global symmetry
 and/or discrete symmetry can be imposed to obtain the
needed texture of mass matrices.

To explain the
universality of the Yukawa interactions
one can also develop  the  scenario
of  Sect. III. Namely, it is possible to generate the
 the Majorana mass matrix of \nuR
via interactions with additional
superheavy leptons.

\vskip 0.8cm

\noindent
{\bf V. Discussion and  conclusions}

1. The most natural scenario of lepton mass and mixing generation
implies that the Dirac mass matrices in the lepton
sector are similar to those in the quark sector.
The difference in the neutrino masses and
probably in mixing follows from
 the Majorana mass
matrix of the right-handed neutrinos.

2. The effect of the see-saw mechanism itself in lepton mixing
can be described by the see-saw matrix or
the see-saw angle (when the task is reduced to the
two-neutrino case).  The see-saw angle is proportional to
the mass hierarchy in the Dirac sector $\epsilon ^D$, rather than
$\sqrt {\epsilon ^D}$. Therefore,  usually the see-saw corrections
to lepton mixing  are small, or are of the same order as  mixing
 induced by the Dirac matrices. However the smallness of the
see-saw angle related to such a proportionality can be systematically
compensated.
At definite conditions,
the see-saw angle can be  larger than the angles
induced by Dirac matrices and
the lepton mixing is determined mainly by the see-saw angle.

3. There are two sets of conditions at which the see-saw
angle dominates and  the lepton mixing is
large.
The first set demands a strong mass hierarchy in
the Majorana sector as well as approximately the same
transformations of the right-handed neutrino components
which diagonalize the Dirac and Majorana mass matrices. The latter
implies the universality of the Yukawa couplings of the
right-handed components.

These  conditions can be realized in  models with
additional superheavy leptons, $N_L$,
where  right-handed neutrinos
acquire Majorana masses via the see-saw mechanism induced
by interactions with $N_L$. The universality of  Yukawa
couplings of \nuR could be explained by the unification of
  \nuL and $N_L$ in one multiplet.
As a result of the extended (cascade) see-saw, the mixing
of the light neutrinos is determined essentially by  mixing of   $N_L$.
The typical mass scale of light neutrinos, ($10^{-3} - 10^{-1}$) eV,
allows us to explain both the solar and atmospheric neutrino
deficits.

4. The second set of conditions demands definite form of the Majorana
mass matrix of right-handed components in the Dirac  basis of neutrinos
(where Dirac mass matrix is diagonal).
Namely, dominance of the nondiagonal elements is needed and
in most natural cases
the Majorana masses of right-handed neutrinos turn out to be degenerate.
Such a nondiagonality of $M$ (or suppression of
the diagonal elements) is a generic feature of models
in which neutrinos have nonzero charges of some  symmetry, $G$.
For this scenario an enhancement of mixing for two neutrinos
is typical. Different $G$-charge prescriptions result in enhancement
of mixing of different components.

5. Future solar neutrino experiments will be able
to confirm (or reject) the solutions based on large lepton
mixing and therefore check (at least partly) the possibilities
considered in this paper.

\vskip 0.8cm

\noindent
{\bf Acknowledgements}

I would like to thank M.~Fukugita and G.~Senjanovi\'c
 for valuable discussions. I am grateful to M.~Kamionkowski
for reading the manuscript and to M.~Best and P.~I.~Krastev for help in
the preparation of the
paper.  Work was supported in part by the National Science Foundation
(grant {\bf NSF \#PHY92-45317}) and the Ambrose Monell Foundation.

\vskip 0.8cm

\noindent
{\bf Appendix. See-saw angle and see-saw matrix.}

In the two-neutrino case, the mass matrix is diagonalized
by the unitary transformation  $S(\theta_i)$ which is determined by
a rotation angle $\theta_i$ :
$$
S(\theta_i) \equiv
\pmatrix{
\cos \theta_i & \sin \theta_i \cr
- \sin \theta_i & \cos \theta_i \cr
}\ .
\eqno(A.1)
$$
Let $\theta_{L}^{D}$ and $\theta_{R}^{D}$ be the angles
of rotations of the left and right-handed neutrino components,
$\nu_{L} = S(\theta_{L}^{D}) \nu_{L}^{D}$ and
$\nu_{R} = S(\theta_{R}^{D}) \nu_{R}^{D}$, which diagonalize the
Dirac mass matrix:
$$
S^{T}(\theta_{L}^{D}) m_{D} S(\theta_{R}^{D}) = m_{D}^{diag}
\equiv diag(m_1, m_2)\ .
\eqno(A.2)
$$
Here $\nu_{L}^{D}$, $\nu_{R}^{D}$ and $m_1$, $m_2$ are the
eigenstates and the eigenvalues of the Dirac mass matrix
correspondingly (subscript $T$ means transponent). Let
\tm be the angle of rotation of the
right-handed neutrino components which diagonalizes the Majorana
mass matrix of right-handed neutrino components,
$\nu_{R} = S(\theta^{M}) \nu_{R}^{M}$:
$$
S^{T}(\theta^{M}) M S(\theta^{M}) = M^{diag}
\equiv diag(M_1, M_2)\ ,
\eqno(A.3)
$$
where $M_i$ are the eigenvalues of the Majorana mass matrix.
Then taking into account that the inverse matrix
$M^{-1}$ is diagonalized by the same transformation (A.3)
we can write the Majorana mass matrix for light neutrino
components in terms of diagonalized
matrices and rotations in the following form:
$$
m^{maj} =
S(\theta_{L}^{D}) m^{diag}_{D} S(\theta^{M} -
\theta_{R}^{D}) (M^{diag})^{-1}
S^{T}(\theta^{M} - \theta_{R}^{D}) m^{diag}_{D} S(\theta_{L}^{D})^T\ .
\eqno(A.4)
$$
This matrix can be diagonalized by transformation (A.1) with
the angle
$$
\theta_{\nu} \equiv \theta_{L}^{D} + \theta_{ss}\ ,
\eqno(A.5)
$$
where \tss  is the angle of rotation
which  diagonalizes according to (A.4) the matrix
$$
m_{ss} \equiv
m^{diag}_{D} S(\theta^{M} -
\theta_{R}^{D}) (M^{diag})^{-1}
S^{T}(\theta^{M} - \theta_{R}^{D}) m^{diag}_{D}\ .
\eqno(A.6)
$$
Now the lepton mixing angle can be written  as
$\theta^{lept} =  \theta_{L}^{D} + \theta_{ss} - \theta_{L}^{l}$;
here $\theta_{L}^{l}$ is the angle of rotation which
diagonalizes the Dirac mass matrix of charged leptons.
At \tss = 0 the form of lepton mixing coincides with that in
quark sector. So, \tss specifies the features of
the see-saw mechanism and we will call it the see-saw angle.

In the general case (three-neutrino mixing), one can introduce the
see-saw matrix, $S_{ss}$, so that the lepton mixing matrix can be
written in the form
$$
S^{lept} = (S_L^l)^+ S^D_L S_{ss}\ ,
\eqno(A.7)
$$
where the transformation $S_{ss}$ diagonalizes the matrix
$$
m_{ss} = m_D ^{diag}S_R^{T} M^{-1} S_R m_D^{diag}\ .
$$

\vskip 0.8cm

\noindent
{\bf References}

\Rf
[1] M.~Gell-Mann, P.~Ramond, R.~Slansky, in {\it Supergravity},
ed.~by F.~van Nieuwenhuizen and
 D.~Freedman (Amsterdam,
North Holland, 1979) 315;
 T.~Yanagida, in Proc.~of the Workshop on the
{\it Unified Theory and Barion Number in the Universe}, eds.~
O.~Sawada and A.~Sugamoto (KEK, Tsukuba) 95 (1979);
R.~N.~Mohapatra and G.~Senjanovi\'c, Phys.~Rev.~Lett.
{\bf 44}, 912 (1980).
\Rf
[2] P.~Langacker, Phys.~Rep.,  {\bf 72}, 185 (1981);
 J.~Vergados, Phys.~Rep., {\bf 133}, 1 (1986);
 S.~M.~Bilenky and S.~T.~Petcov,
Rev.~Mod.~Phys., {\bf 59} 671 (1987)
 J.~W.~F.~Valle, Prog.~Part.~Nucl. Phys.~{\bf 26},
91 (1991).
\Rf
[3] H.~Nishiura, C.~W.~Kim, J.~Kim, Phys. Rev.~D {\bf 31}, 2288
(1985).
\Rf
[4] R.~Johnson, S.~Ranfone and J.~Schechter, Phys.~Lett.~B {\bf 179},
355 (1986).
\Rf
[5] A.~Bottino, C.~W.~Kim, H.~Nishiura, W.~K.~Sze, Phys.~Rev.~D {\bf 34}, 862
(1986).
\Rf
[6] H.~Harari and Y.~Nir, Nucl.~Phys.~{\bf B292}, 251 (1987).
\Rf
[7] G.~C.~Branco, W.~Grimus, L.~Lavoura, Nucl.~Phys.~B {\bf 312}, 492
(1989).
\Rf
[8] C.~H.~Albright, Phys.~Rev.~D {\bf 43}, R3595 (1991);
Phys.~Rev.~D {\bf 45}, R725 (1992).
\Rf
[9] G.~Lazaridis and Q.~Shafi, Nucl.~Phys.~B {\bf 350}, 179
(1991).
\Rf
[10] S.~A.~Bludman, D.~C.~Kennedy and P.~Langacker, Nucl.~Phys.,
{\bf B373}, 498 (1992).
\Rf
[11] K.~S.~Babu and Q.~Shafi, Phys.~Lett.~B {\bf 294}, 235 (1992).
\Rf
[12] S.~Dimopoulos, L.~Hall and S.~Raby, Phys.~Rev.~Lett.~
{\bf 68}, 1984 (1992), Phys.~Rev.~D {\bf 45},
 4192 (1992).
\Rf
[13] M.~Fukugita, M.~Tanimoto and T.~Yanagida, Prog.~Theor.~Phys.~
{\bf 89}, 263 (1993).
\Rf
[14] A.~Yu.~Smirnov, Proc.~of the {\it Joint Int.~LP-HEP Symposium}
(Geneva, 1991),
 S.~Hegarty, K.~Potter and E.~Quercigh, v.1, 623 (1992).
\Rf
[15] R.~Davis, Talk given at the {\it International
Symposium on Neutrino Astrophysics},
\break Takayama/Kamioka (October 1992)
 K.~S.~Hirata et al., Phys.~Rev.~Lett.~{\bf 66}, 9 (1991);
Phys.~Rev.~D {\bf 44}, 2241 (1991);
 SAGE: A.~I.~Abazov et al.~Phys.~Rev.~Lett., {\bf 67},
 3332 (1991);
 GALLEX: P.~Anselmann et al., Phys.~Lett.~B  {\bf 285},
376 (1992).
\Rf
[16] J.~N.~Bahcall and M.~H.~Pinsonneault, Rev.~Mod.~Phys.~{\bf 64}, 885
(1992); S.~Turck-Chieze
 et al.~Astroph.~J.~{\bf 335}, 415 (1988); I.-J.~Sackmann et al.,
Astroph.~J.~{\bf 360}, 727 (1990).
\Rf
[17] V.~N.~Gribov and B.~M.~Pontecorvo, Phys.~Lett. {\bf 28}, 493 (1967);
J.~N.~Bahcall and
 S.~C.~Frautschi, Phys.~Lett.~B {\bf 29}, 623 (1969);
V.~Barger, R.~J.~N.~Phillips, and K.~Whisnant,
 Phys.~Rev.~D {\bf 24}, 538
(1981);
S.~L.~Glashow and L.~M.~Krauss, Phys.~Lett.~B {\bf 190}, 199
 (1987).
\Rf
[18] V.~Barger, R.~J.~N.~Phillips and K.~Whisnant, Phys.~Rev.~D {\bf 43},
1110 (1991);
A.~Acker,
 S.~Pakvasa and J.~Pantaleone, Phys.~Rev.~D {\bf 43}, 1754
(1991); P.~I.~Krastev and
 S.~T.~Petcov, Phys.~Lett.~B {\bf 285}, 85 (1992).
\Rf
[19] S.~P.~Mikheyev and A.~Yu.Smirnov, Sov.~J.~Nucl.~Phys.~{\bf 42}, 913
(1986); Prog.~Part.~Nucl.
 Phys., {\bf 23}, 41 (1989);
 L.~Wolfenstein, Phys.~Rev.~
 D {\bf 17}, 2369 (1978), ibidem, D {\bf 20}, 2634 (1979);
 T.~K.~Kuo and J.~Pantaleone,
Rev.~Mod.~Phys.~{\bf 61}, 937 (1989).
\Rf
[20] P.~Anselmann et al., Phys.~Lett.~B {\bf 285}, 389 (1992);
 S.~A.~Bludman, D.~C.~Kennedy and P.~G.~Langacker, Phys.~Rev.~
D {\bf 45},
1810 (1992); S.~A.~Bludman, N.~Hata, D.~C.~Kennedy and
P.~G.~Langacker, Pennsylvania preprint UPR-0516T;
 X.~Shi and D.~N.~Schramm, Phys.~Lett.~B {\bf 283}, 305 (1992);
X.~Shi,  D.~N.~Schramm, and
 J.~N.~Bahcall, Phys.~Rev.~Lett.~{\bf 69}, 717 (1992);
 J.~M.~Gelb, W.~Kwong and S.~P.~Rosen, Phys.~Rev.~Lett.~
{\bf 69}, 1846 (1992);
 W.~Kwong and S.~P.~Rosen, Phys.~Rev.~Lett.~{\bf 68},
748 (1992);
 A.~Yu.~Smirnov, ICTP preprint IC/92/429;
 P.~I.~Krastev and S.~T.~Petcov, Phys.~Lett.~B {\bf 299},
99 (1993);
 L.~M.~Krauss, E.~Gates and M.~White, Phys.~Lett.~B {\bf 298},
94 (1993),
 Phys.~Phys.~D {\bf 46}, 1263 (1992),
Phys.~Rev.~Lett., {\bf 70}, 375 (1993).
\Rf
[21] K.~S.~Hirata et al., Phys.~Lett.~B {\bf 280},146 (1992);
 R.~Becker-Szendy et al.~Phys.~Rev.~D {\bf 46},
3720 (1992).
\Rf
[22] H.~Fritzsch, Phys.~Lett.~B {\bf 70}, 436 (1977); ibidem {\bf 73} 317
(1978).
\Rf
[23] A.~Yu.~Smirnov, D.~N.~Spergel and J.~N.~Bahcall,
Preprint IASSNS-AST 93/15.

\end